# Origins of large enhancement in electromechanical coupling for nonpolar directions in ferroelectric BaTiO$_3$

A. Pramanick,[1] S. O. Diallo,[2] O. Delaire,[3] S. Calder,[2] A. D. Christianson,[2] X.-L. Wang[4] and J. A. Fernandez-Baca[2]

[1]*Chemical and Engineering Materials Science Division, Oak Ridge National Laboratory, Oak Ridge, TN 37831 USA*
[2]*Quantum Condensed Matter Division, Oak Ridge National Laboratory, Oak Ridge, TN 37831 USA*
[3]*Materials Science and Technology Division, Oak Ridge National Laboratory, Oak Ridge, TN 37831 USA*
[4]*Department of Physics and Materials Science, City University of Hong Kong, Hong Kong SAR, PRC*

**The origins of enhanced piezoelectric coupling along nonpolar crystallographic directions in ferroelectric BaTiO$_3$ are investigated using *in situ* neutron spectroscopy. It is observed that an electric field applied away from the equilibrium polarization direction causes changes in the phonon spectra that lead to an increase in the interaction between the transverse acoustic and transverse optic branches (TA-TO) near the Brillouin zone center. This provides a direct lattice dynamics mechanism for enhanced electromechanical coupling, and could act as a guide for designing improved piezoelectric materials.**

Piezoelectric materials are attractive for their electromechanical coupling properties and are widely used in applications ranging from diagnostic medical imaging, to sonar, precision actuators, telecommunications, and energy harvesting.[1] The electromechanical properties of a material are often characterized by the piezoelectric coefficient *d*, which is the ratio between the resulting mechanical strain and the applied electric field which causes it. In certain ferroelectric single crystals and grain-oriented ceramics, *d* is dramatically enhanced for specific applied electric field directions that are different from the spontaneous polarization vector of the crystal lattice.[2-4] A detailed understanding of the origins for such enhancement has so far remained elusive. For example, in relaxor ferroelectric crystals of $Pb(Zn_{1/3}Nb_{2/3})O_3$-$PbTiO_3$ with compositions close to the morphotropic phase boundary (MPB), a maximum $d_{33}$ ~2500 pm/V is observed for applied electric fields along <001>; while, $d_{33}$ ~100 pm/V when the electric field is applied parallel to the polar <111> direction.[2] One possible mechanism put forward for such giant enhancement in piezoelectric properties was the reorientation of the polar nanoregions (PNR) within a compositionally disordered microstructure.[5] It was postulated that the reorientation of the PNRs creates a phase instability which in turn could lead to a greater electric-field-induced strain response.[6] However, subsequent observations showed that PNRs are not essential for this effect and similar effects can also be observed in normal ferroelectrics with no presence of PNRs or compositional disorder.[3-4] It was therefore realized that there should be an intrinsic crystallographic phenomenon responsible for the enhancement of electromechanical properties in ferroelectric crystals along nonpolar crystallographic directions.

Most well-known ferroelectrics are perovskite compounds with the general formula of $ABO_3$. The perovskite crystal structure is characterized by the archetype ferroelectric $BaTiO_3$, in which the Ba ions occupy the cell corners, O ions occupy the face centers and the Ti ion is closest to the body center of the unit cell. Upon cooling below the ferroelectric Curie temperature $T_C$, $BaTiO_3$ undergoes a sequence of structural transitions from the cubic-to-tetragonal-to-orthorhombic-to-rhombohedral phases at successive lower temperatures. Within the framework of displacive phase transitions, the Ti ion is considered to be displaced from the center of the unit cell along the polarization direction of <001>, <011> and <111> within the tetragonal, orthorhombic and rhombohedral phases, respectively. The polarization vector can however be rotated away from its equilibrium direction upon application of an electrical field along certain non-polar directions. Both first-principles and phenomenological calculations indicate that the energy surface for rotation of the polarization vector is flatter for specific directions of applied electric fields,[7] and that the rotation of the polar vector is related to an increased shear of the crystal lattice.[8-10] However, the microscopic mechanism through which polarization rotation might be related to an increased lattice deformation is unknown. Moreover, the standard model of polarization rotation for enhanced electromechanical properties raises additional concerns. For example, it is debatable whether the ferroelectric distortion in $ABO_3$ compounds is simply caused by an off-centering of the B cation within the oxygen octahedra.[11] Moreover, such a model cannot explain experimentally observed *irreversible* electric-field-induced structural changes in the crystal symmetry.[3,12] Therefore, a fundamental understanding of the effects of a nonpolar electric field is required.

The central role of lattice dynamics towards the electromechanical properties of crystalline materials was originally recognized by Cochran.[13] From an atomistic viewpoint, the piezoelectric properties of crystalline solids result from a coupling between the specific vibration modes related to the elastic deformation of the lattice and to the generation of an electric dipole, respectively. In ferroelectric crystals, the former is connected to the long-wavelength acoustic phonons and the latter is connected to a soft transverse optic phonon mode near the Brillouin zone center.[14] Although an interaction between the transverse acoustic (TA) and the transverse optic (TO) modes exists in all piezoelectric ionic crystals, it is stronger when the frequencies of the two modes approach each other. This is observed in ferroelectric crystals for which the frequency of the transverse optic mode is low near the zone center or ZC ("soft" mode) and therefore partially overlaps with the acoustic mode, consequently leading to high piezoelectric properties.[14-16] Therefore the origins of enhanced piezoelectric properties in ferroelectric crystals for applied nonpolar electric fields can be examined from the induced changes in the TA and TO modes near the ZC.

$BaTiO_3$ was selected as the prototype material to undertake such a study. Particular phonon modes were investigated that could elucidate induced lattice instabilities that are related to higher piezoelectric response under application of electric fields along nonpolar directions.[10]

Specifically, static electric fields of increasing magnitudes were applied along the nonpolar [110]$_{pc}$ direction of a tetragonal BaTiO$_3$ crystal and electric-field-induced changes were examined for two vibrational modes that are correlated to the ferroelectric instability at the $\Gamma$ point (0,0,0): namely the transverse acoustic (TA) $\Lambda_3$ mode extending from the $\Gamma$ point to the *R* point (0.5,0.5,0.5) and the transverse optic (TO) $\Gamma_{25}$ mode which is the ferroelectric soft mode at the ZC.[17,18] We show that the application of a nonpolar electric field leads to an upward shift in energy of the TA branch, thereby bringing it closer to the TO ferroelectric soft mode. An increased overlap between these modes leads to higher interaction between the TA and TO phonons near the ZC and therefore provides a direct microscopic mechanism for enhanced electromechanical coupling. The origin of an electric-field-induced shift in the TA dispersion branch is subsequently examined within the context of an order-disorder nature of structural distortions in BaTiO$_3$.

The inelastic neutron scattering experiments were performed on the triple-axis spectrometer HB-3 at the High Flux isotope Reactor at the Oak Ridge National Laboratory. The inelastic spectrometer was operated with a fixed final energy of E$_f$ = 14.7 meV. Pyrolytic Graphite (PG) crystals were used as monochromator and analyzer. The spectrometer was first set to tight collimation mode of 48'-20'-20'-70' but was subsequently relaxed to 48'-40'-40'-120' to improve the signal-to-noise ratio. Two single crystals of BaTiO$_3$ obtained from different batches of dimensions 10 mm × 1 mm × 1mm were obtained from MTI Croporation, CA, USA. The crystals were identically cut with faces normal to <110> and <001>pseudocubic directions, with 110 being the normal to the face with the larger surface area. The principal result, that the TA branch hardens, is observed in both the crystals. Similar behavior in both samples supports that the observations reported here are intrinsic to BaTiO3. For the neutron scattering experiments, the crystals were oriented in the HHL plane and electric field was applied normal to the 110 crystal face. Both constant-wavevector (Q) and constant-energy (E) scans were employed. The TA and TO phonon peaks for the respective $\Lambda_3$ and $\Gamma_{25}$ modes were measured by scanning along the [1+ζ,1+ζ,-1+ζ] direction from the $(11\bar{1})$ Bragg peak, where ζζζ defines the reduced wave vector **q.** We not that the 111 direction is common to all tetragonal domain variants and therefore provides a practical advantage in terms of measuring phonons propagating along the ζζζ wave vector**.** All measurements were carried out under ambient conditions at room temperature.

Figure 1 shows the phonon spectra measured at constant energy values of E = 5 meV and E =7.5 meV, for the virgin state and under applied electric fields of V = 1.25, 1.5 and 1.8 kV/mm. With increasing electric field from 1.25 kV/mm to 1.8 kV/mm, the peak profiles for TA and TO modes are observed to change - a progressive dampening of the TO mode is clearly observed. The phonon peaks associated with the TA $\Lambda_3$ mode and the TO $\Gamma_{25}$ soft mode are fitted to Lorentzian profiles to extract the peak parameters. The TA dispersion branch over an extended q range is plotted in Figure 2(a) for the virgin state and for V=1.5 kV/mm (for E=5 meV, V=1.25 kV/mm is used for calculation due to increased overlap between the modes at higher fields). The shift in the peak positions associated with the TA $\Lambda_3$ mode in the constant-E scans, for the two different states of electric fields, can be observed from Figure 2(b). The shift of the peak positions to a lower q under the applied electric field corresponds to a stiffening of this dispersion branch. Additionally, there is an anomalous broadening of both the TA and TO modes, particularly at higher electric fields. Because of the shift in the TA branch, a greater overlap of the two modes occurs upon electric field application. We also note that the shift in the TA dispersion is not proportional in q. This is apparent from Figure 2(c) in which the shift in q under field application is plotted as a percentage of the absolute q value at zero field. The relative shift in *q* shows a dip for *q* ~ 0.12. In addition to peak shifts, the total intensity of the TA+TO phonon spectra also changed with increasing electric field as shown in Figure 2(d). The intensity change could reflect a more diffuse contribution of the modes to the dynamical structure factor, S(**Q**,E), or a quantum-mechanical interference between TA and TO modes.

The above features related to anomalous peak profile and dispersion of phonon modes at finite $q$ values can be expected as a consequence of coupling between the original eigen states related to the TA and TO modes. As the two phonon modes approach each other to have the same value of the wave vector $q$, a quantum mechanical interference between the two competing scattering channels occurs.[17,18] If the modes are out-of-phase with respect to each other, addition of the respective phonon peaks will result in a decrease of the total TA+TO peak intensity as is observed in Figure 2(d), and redistribution of the spectral weight in the dynamical structure factor, S(**Q**,E). Naturally, such coupling between an optic mode at $q \approx 0$ and an acoustic mode dispersed in $q$ will be largest near the zone center and will decrease with increasing $q$. The effects of a $q$ dependent TA-TO mixing can be observed from anomalous dispersion of acoustic phonons, such as shown in Figure 1 of Ref. [19] by Shirane *et al*. for $PbTiO_3$,[19] similar to a "dip" as observed in Figure 2(c). As compared to $PbTiO_3$, additional anharmonic interference between TA and TO modes in $BaTiO_3$ is caused by a broadening of either mode and consequent overlap of their excitation spectra.[17,18] Therefore, the currently observed anomalous features in the phonon dispersions and peak profiles are consistent with the well-known effects of TA-TO mode coupling in ferroelectric crystals. Most remarkably, the present observations show that TA-TO coupling is enhanced when an electric field is applied along a nonpolar crystallographic direction.

The increased mode coupling essentially results from a hardening of the TA mode which brings it closer to the TO mode. It is worthwhile to mention here that calculations based on Landau-Devonshire-Ginzburg (LGD) type phenomenological theories have indicated that the enhanced piezoelectric properties along nonpolar directions in ferroelectric crystals could be related to a softening of the transverse dielectric coefficient.[9,20] Unlike the *dielectric* coefficient, the current results indicate no such role for the transverse *elastic* constant. On the contrary, a hardening of the TA mode would be consistent with an increase in the elastic constant under a nonpolar electrical bias. In the following, we examine why a hardening of the TA mode could occur in $BaTiO_3$.

Lattice distortions in $BaTiO_3$ have been described to have both displacive and order-disorder characteristics.[13,21] In the order-disorder case, $BaTiO_3$ can be described to be intrinsically disordered with the Ti ions displaced along either of the <111> directions. Depending on how the localized distortions are correlated along the different crystallographic directions, the macroscopic symmetry could be either cubic, tetragonal, rhombohedral or orthorhombic.[21] In $BaTiO_3$, the displacement of the Ti ions against the oxygen ions along Ti-O-Ti chains also dominates the ferroelectric instability at the $\Gamma$ point.[22] Consequently, the dispersion of the TA $\Lambda_3$ mode is dependent on the degree of correlation between the different localized displacements of Ti - a steeper dispersion is expected from a strongly-correlated Ti displacements along the Ti-O-Ti chains.[22,23] It is therefore conceivable that when an electrical bias is applied along the $[110]_{pc}$ direction, the correlation among the localized Ti displacements along a specific crystallographic axis is increased and is responsible for "hardening" of the TA dispersion branch .

In order to further elaborate on the nature of propagation of the acoustic phonons, we measured the distribution of the TA mode near the point A in Figure 1, for different values of electric fields, by scanning across E with a constant *q* value. We chose point A in order to avoid tunneling effects from Bragg scattering close to the ZC and the region of *q* where there is a large overlap of the peaks for the TA and the TO modes. The FWHMs for the phonon peaks at different applied field amplitudes are obtained from Lorentzian fits, as shown in Figure 3. There is a steady increase in the FWHM with electric field amplitude implying a decrease in the phonon lifetime. We identify the following two causes for this effect. It is intuitively understood that when two phonon modes overlap with each other, a single one phonon excitation of the lattice will decay into multiple correlated final phonon states; this will cause an accelerated decay of the original phonon excitation and consequently cause a decrease in the phonon lifetimes. Alternately, phonon lifetimes could be influenced by changes in short range ordering of localized atomic displacements, as the phonons are scattered at the interfaces between the differently ordered regions.[17,18] The later effect is less likely since more correlated off-centerings of Ti atoms would lead to a more ordered structure and scattering from interfaces between disordered regions. We therefore propose that the decrease in phonon lifetimes is also a result of increased coupling between the TA-TO modes upon electric field application.

Finally, we observe that the shift in the TA phonon dispersion persists even subsequent to the withdrawal of the electric field (Figure 4). This could indicate a permanent change in the atomic arrangements. An irreversible electric-field-induced lattice distortion in $BaTiO_3$, for fields applied along nonpolar crystallographic directions, is in line with previous observations of permanent changes in crystal symmetry in $BaTiO_3$ single crystals.[3,12] While the intrinsic disorder for atomic displacements in $BaTiO_3$ has been described as essentially a dynamic phenomenon,[18] it is also possible that an ordered phase is induced upon application of an electrical bias.[7] The latter view is supported from the present results. Additional insights in this area can be gained from diffused scattering measurements under applied electric fields.

The current study shows the significance of *in situ* neutron spectroscopy characterization of phonons towards identifying the potential microscopic phenomena that can lead to giant electromechanical properties in solids. The key phenomenon observed here is an increase in TA-TO coupling with electric field application, which likely originates from changes in chain-like correlations among individual atomic displacements. A clearer understanding of the nature of intermixing among the different phonon modes is expected from detailed line width analyses of phonon spectra measured under applied electric fields. Other types of correlated structural changes in perovskite compounds include cooperative tilts of oxygen octahedra[24,25] and changes in compositional disorder among the different atomic sites.[6] Understanding how these structural changes can affect the acoustic and optic phonon modes *during* electric field application will be critical for development of emerging relaxors and non-lead ferroelectrics with high electromechanical properties.

**Ackowledgments:** Research conducted at ORNL's High Flux Isotope Reactor was sponsored by the Scientific User Facilities Division, Office of Basic Energy Sciences, US Department of Energy. Funding from a Laboratory Directed Research and Development Fund of Oak Ridge National Laboratory is acknowledged. The authors also acknowledge technical assistance from Christopher M. Redmon and Daniel Maierhafer on High Voltage experimental setup.

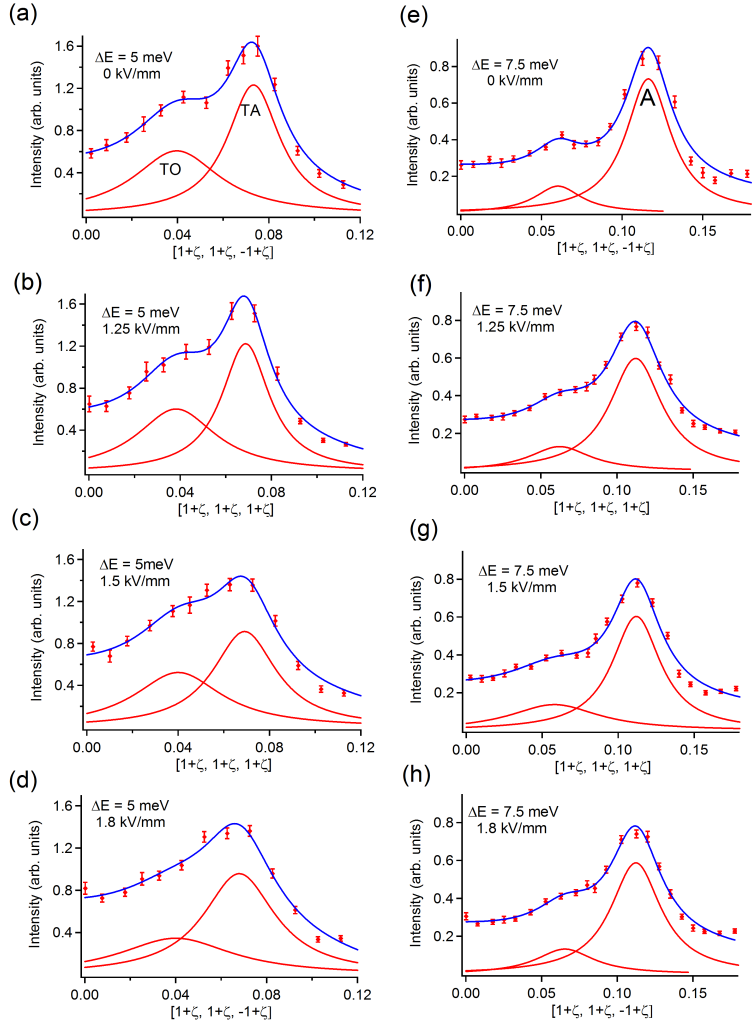

Figure 1: Constant-energy scans along the [1+ζ,1+ζ,-1+ζ] direction showing the peak positions for the $\Lambda_3$ TA and $\Gamma_{25}$ TO phonon modes: (a)-(d) scans at ΔE = 5 meV for fields of V = 0, 1.25, 1.5 and 1.8 kV/mm, respectively; (e)-(h) scans at ΔE = 7.5 meV for fields of V = 0, 1.25, 1.5 and 1.8 kV/mm, respectively.

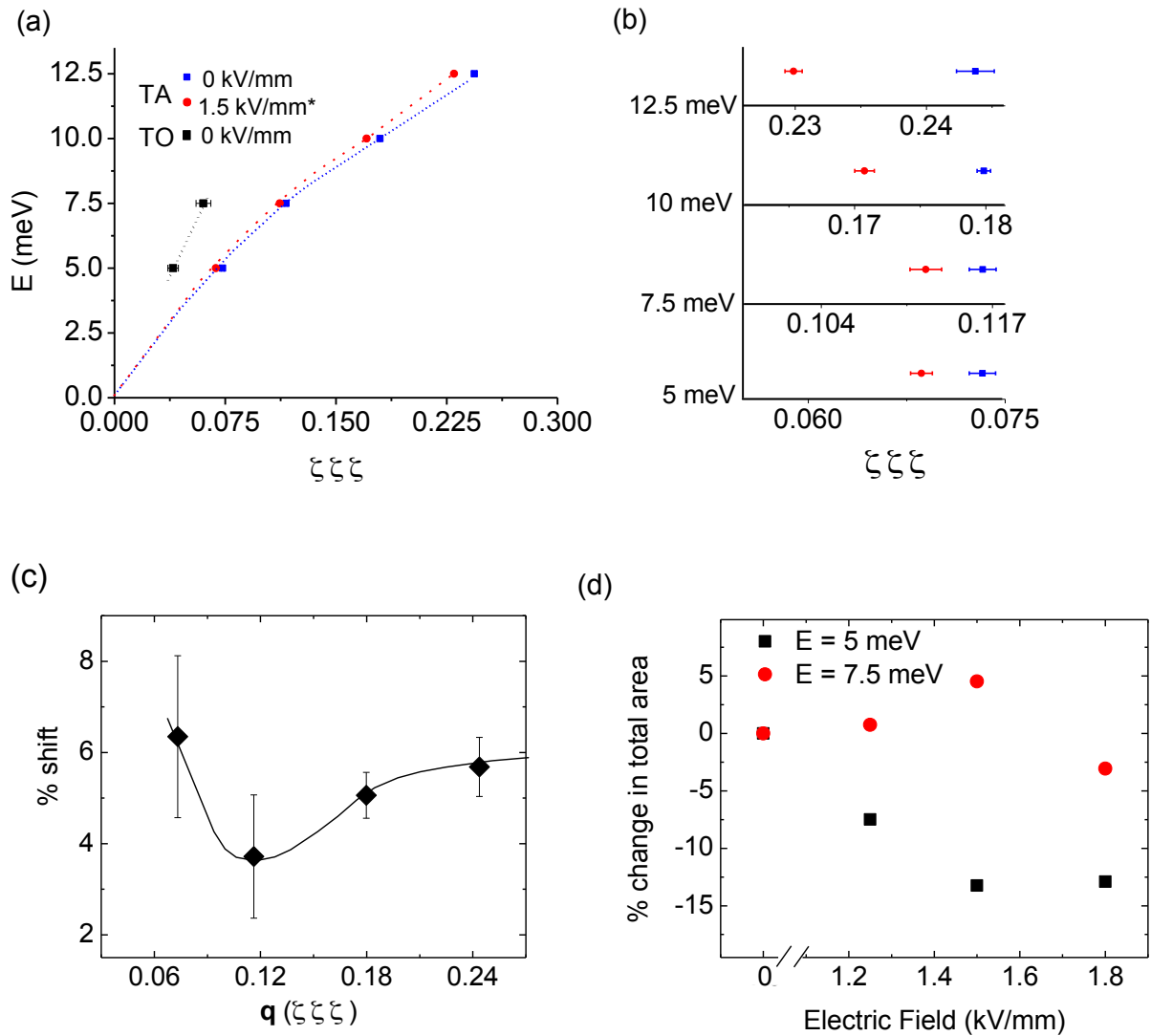

Figure 2: Electric-field-induced changes in phonon spectra: (a) the $\Lambda_3$ TA dispersion branch over an extended q range for zero field and for V=1.5 kV/mm; the $\Gamma_{25}$ TO dispersion near the ZC is shown for zero field (b) the shift in TA peak positions can be observed at each ΔE (*the peak position for E=5 meV was calculated for V=1.25 kV/mm, as at higher applied fields the TA and TO modes strongly overlap each other and separation between the two becomes difficult); (c) the shifts in *q* for TA peaks are plotted as a percentage of the absolute *q* value at zero field; (d) percentage change in total peak area (TA+TO) as a function of applied electric field magnitude.

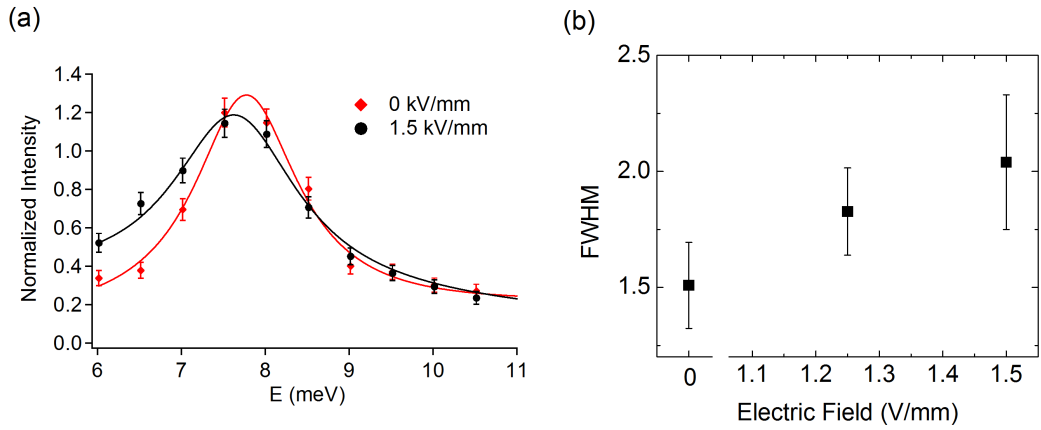

Figure 3: Effect of applied electric field on phonon line width: (a) $\Lambda_3$ TA phonon measured with constant-Q scan near the point A in Figure 1, at zero field and V = 1.5 kV/mm; (b) field-dependence of TA peak width presented as FWHM

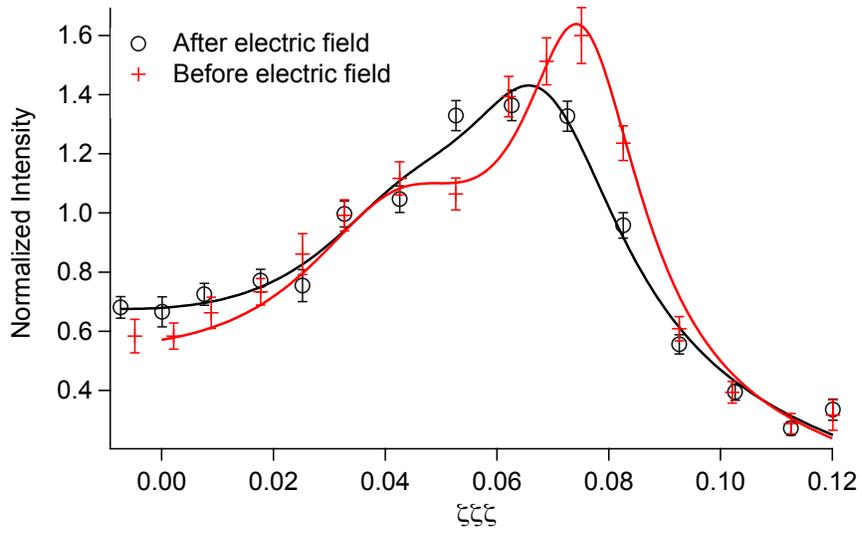

Figure 4: Irreversible change in the phonon spectra before and subsequent to the application of electric fields